\documentclass[aps,prl,superscriptaddress,twocolumn,english]{revtex4-1}
\usepackage[T1]{fontenc}
\usepackage[latin9]{inputenc}
\usepackage{color}
\usepackage{verbatim}
\usepackage{amstext}
\usepackage{amssymb}
\usepackage{graphicx}
\usepackage{bm}
\makeatletter
\usepackage{amsmath}
\usepackage{comment}
\newcommand{\lyxdot}{.}
\makeatother

\usepackage{babel}
\begin{document}

\title{Theory of magnetic response in finite {two-dimensional}  superconductors}

\author{F. Sebasti\'an Bergeret}

\affiliation{Centro de F\'isica de Materiales (CFM-MPC), Centro Mixto CSIC-UPV/EHU,
Manuel de Lardizabal 4, E-20018 San Sebasti\'an, Spain}
\affiliation{Donostia International Physics Center (DIPC), Manuel de Lardizabal
5, E-20018 San Sebasti\'an, Spain}

\author{Ilya V. Tokatly}

\affiliation{Nano-Bio Spectroscopy group, Dpto. F\'isica de Materiales, Universidad
del Pa\'is Vasco, Av. Tolosa 72, E-20018 San Sebasti\'an, Spain }
\affiliation{IKERBASQUE, Basque Foundation for Science, E-48011 Bilbao, Spain}
\affiliation{Donostia International Physics Center (DIPC), Manuel de Lardizabal
5, E-20018 San Sebasti\'an, Spain}

\begin{abstract}
We present a theory of magnetic response in a finite-size  two-dimensional superconductors with Rashba spin-orbit coupling.  The interplay between the latter and an in-plane Zeeman field leads on the one hand to an out-of-plane spin polarization which accumulates at the edges of the sample over the superconducting coherence length, and  on  the other hand, to circulating supercurrents decaying away from the edge over a macroscopic scale. { In a long finite stripe of width $W$ both, the spin polarization and the currents, contribute to 
the total magnetic moment induced at the stripe ends. {These two  contributions   scale with powers of $W$ such that for sufficiently large samples it can be detected by current  magnetometry techniques.} }
\end{abstract}
\maketitle
Superconductivity  in two-dimensional (2D) and quasi-2D systems  has been attracting a great deal of attention over past decades \cite{uchihashi2016two,saito2017highly}.
Examples of such systems range from ultra-thin metallic films,
heavy fermion superlattices, and interfacial superconductors to
atomic layers of metal dichalcogenides, and organic conductors. 

Most 2D superconductors exhibit large spin-orbit coupling (SOC)
because of broken space inversion symmetry. In this regard
two types of 2D superconductors can be distinguished: Those
exhibiting SOC of Rashba-type due to a broken up-down (out-of-plane) mirror symmetry, denoted here as Rashba superconductors, 
and those,  in which a 2D in-plane inversion symmetry is broken due to a non-centrosymmetric crystal structure. {The latter are} exemplified by 2D transition metal dichalcogenides \cite{saito2016superconductivity,lu2015evidence}.
To the first group, on which we focus here, belong for example ultra-thin superconducting metallic films \cite{gruznev2014strategy,sekihara2013two,menard2017two}. 

Over the last years Rashba superconductors have been intensively studied as paradigmatic systems where pair correlations coexist with strong intrinsic SOC \cite{edel1989characteristics,edelstein1995magnetoelectric,edelstein1996ginzburg,yip2002two,Gorkov2001,frigeri2004spin,edelstein2008anomalous,agterberg2007magnetic,dimitrova2003phase,dimitrova2007theory,Pershoguba2015,Malshukov2016,Konschelle2015}. Because of  the interplay between SOC and a Zeeman field they demonstrate highly unusual properties, such as, the appearance of an inhomogeneous superconducting phase \cite{dimitrova2007theory,agterberg2007magnetic}, { magnetoelectric effects \cite{edelstein1995magnetoelectric,yip2002two,Konschelle2015}, and anisotropic magnetic susceptibility \cite{Gorkov2001}.  With few exceptions, as for example Refs. \cite{edelstein2003triplet,Pershoguba2015,Malshukov2016},  most of these  works  focused on infinite 2D systems.} 

\begin{figure}
\begin{minipage}[t]{1\columnwidth}%
\includegraphics[scale=0.3]{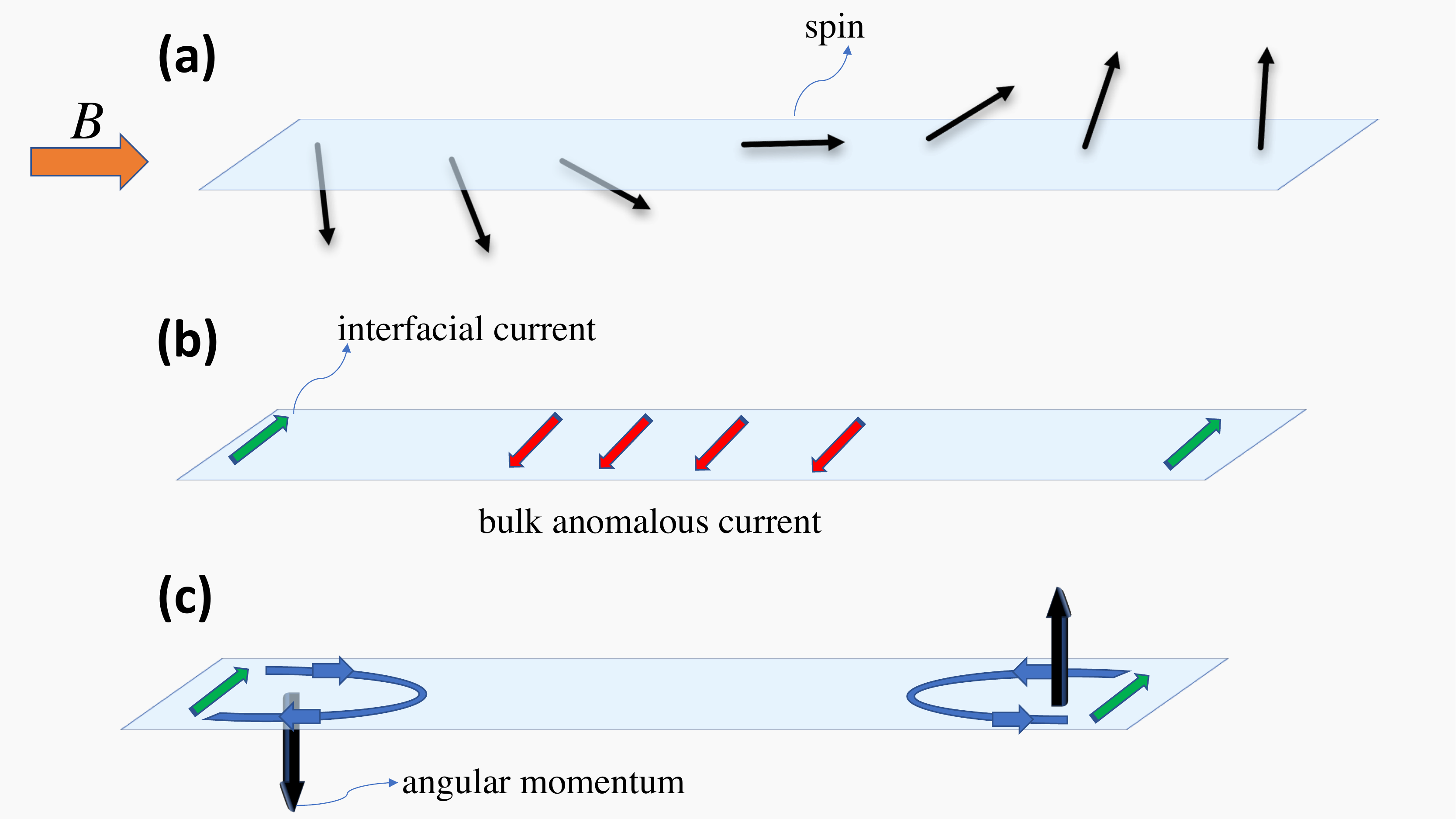}%
\end{minipage}
\caption{Schematically description of the magnetic response in finite size superconductors in the  presence of  an in-plane field $B$. (a) Black arrows represents the deviation of the  spin density, $\delta S$,  from the Pauli spin.  Because of the SOC and the finite size of the sample a transverse component of $\delta S$ is generated.  (b) The spin-charge coupling due to the SOC induces bulk (red arrows) and edge (green arrows)  charge supercurrents. (c) Due to the finite size of the sample the edge currents flow in close loops (blue), inducing  an out-of-plane angular  momentum (black arrows).  \label{fig:Spin-density} }
\end{figure}

In this letter we demonstrate that finite size effects  drastically modify  the  magnetic response of Rashba superconductors leading to hitherto unknown  phenomena. Our main findings are the following: (i)  In a  response to a Zeeman field the system  exhibits a spin texture (Fig. \ref{fig:Spin-density}a) with a transverse component of the spin localized near the edge on the scale of superconducting coherence length. (ii) Because of the spin-charge coupling mediated by the SOC, a non-homogenous charge current appears in the system with a spatial distribution that depends on the direction of the applied field and geometry of the system (Fig. \ref{fig:Spin-density}b); (iii) In particular, for  a finite stripe oriented along the field,  macroscopic currents loops appear at the stripe ends (Fig. \ref{fig:Spin-density}c).  Both, the transverse spin and the edge currents contribute to the total magnetic moment  which  can be detected by state-of-the-art magnetometry techniques.   

These findings can be qualitatively understood recalling  the concepts of spin currents and spin galvanic effect (see Fig. \ref{fig:Spin-density}). 
  The key feature of 2D materials without   up-down mirror symmetry   is the Rashba SOC, $H_R=\alpha ({\bf e}_z\times{\bf v})\cdot{\bm\sigma}$. Here ${\bf e}_z$ is a vector normal to the 2D plane, ${\bf v}={\bf p}/2m$ is the quasiparticle velocity, $m$ its  effective mass, ${\bm\sigma}$ is the vector of Pauli matrices, and $\alpha$ is the SOC constant \footnote{In our notation $\alpha$ has dimensions of momentum and it is proportional to  the usual Rashba constant $\alpha_R=\alpha/m$. Throughout the article we choose the $z$-axis as the axis perpendicular to the superconductor plane.}. The SOC acts as an effective ${\bf p}$-dependent spin splitting field. Let us  assume that the system is subject to an external Zeeman field ${\bf B}$, and for some reason the induced spin polarization ${\bf S}$ differs from  the equilibrium Pauli response $\chi_P{\bf B}$, where $\chi_P$ is the Pauli paramagnetic polarizability.  Then the excess spin $\delta{\bf S}={\bf S}-\chi_P{\bf B}$ will experience an inhomogeneous precession in the effective Rashba field, generating a momentum anisotropy of the density matrix. In the presence of  disorder the precession rate $\hat{R}=i[H_R,\delta{\bf S}\cdot\bm\sigma]$ is balanced by the momentum relaxation, which results in a steady spin current in the bulk of the system ${\cal J}_{bulk,k}^a=-\tau{\rm tr}\langle v_k\sigma^a\hat{R}\rangle=\alpha D(\delta S^{z}\delta_{ka}-\delta S^{k}\delta_{az})$, where $\tau$ is the momentum relaxation time, and $D=\tau v_{F}^2/{2}$ is the diffusion coefficient.  
Under equilibrium conditions $\delta{\bf S}=0$ in normal systems, but in superconductors  pair correlations modify the Pauli response  leading to a finite $\delta {\bf S}$ \cite{abrikosov1962spin,Gorkov2001}. 
This leads to finite equilibrium spin-currents in Rashba superconductors generated by the Zeeman field.  For example, a field  applied in $x$-direction in a bulk superconductor produces a spin-current with an out-of-plane polarization, ${\cal J}_{bulk,x}^{z}=-\alpha D\delta S^{x}$. Due to the spin-Hall magneto-electric coupling in Rashba materials the  bulk spin-current generates a transverse charge current according to  $j_{bulk,y}\propto\alpha^{2}{\cal J}_{bulk,x}^{z}=\alpha^3 D\delta S^{x}$, which is nothing, but the anomalous supercurrent well known for bulk superconductors with SOC \cite{edel1989characteristics,yip2002two,dimitrova2007theory}.

In a finite system currents must vanish at the edges of the sample. This condition can be fulfilled only if the distribution of the excess spin $\delta {\bf S}({\bf r})$ is inhomogeneous near the edge, so that the diffusion spin-current ${\cal J}_{diff,k}^{a}=-D\partial_k\delta S^a$ compensates the bulk contribution.
For concreteness, if we assume a boundary with vacuum at $x=0$, the zero spin-current condition for a field applied in $x$-direction reads:
$D\partial_{x}\delta S^{z}={\cal J}_{bulk,x}^{z}$, which implies
that a finite component $\delta S^{z}(x)$ transverse to the field
is induced at the edges of the sample. In this case the spin density
exhibits a  texture as sketched in Fig. \ref{fig:Spin-density}a \cite{tokatly2019correspondence}. In the presence of SOC both the edge and the bulk spin-currents are converted into a charge current  flowing parallel to the boundary, via the spin-galvanic effect, Fig.~\ref{fig:Spin-density}b. In a realistic finite system currents must
vanish at all edges. The anomalous charge currents at the boundaries should then be compensated by supercurrents which stem from a gradient  of the superconducting
phase. As a consequence, in a stripe geometry, an in-plane field  induces current loops at the edges as shown in Fig. \ref{fig:Spin-density}c.
The magnetic moment induced by this currents and by the transverse  spin can in principle be measured to directly detect the effects we predict here.  In the rest of the paper we provide a quantitative derivation of  these effects, calculate the induced magnetic moment, and propose materials in which our predictions can be verified.

Specifically, we consider  a 2D disordered superconductor with Rashba SOC. We assume that the Fermi energy corresponds to the largest energy scale, so that spectral and transport properties can be accurately described by the quasiclassical Green's functions (GFs)
\cite{Eilenberger1968,larkin1969quasiclassical}. In the diffusive
limit these functions are isotropic in momentum and they obey the Usadel equation which in the presence of a Zeeman field
and Rashba SOC reads \cite{Bergeret2005,Bergeret2013,Bergeret2014}:
\begin{equation}
D\tilde{\nabla}_{k}\left(\check{g}\tilde{\nabla}_{k}\check{g}\right)-\left[\left(\omega+i{\bf h}\cdot\bm\sigma\right)\tau_{3}+\Delta\tau_{2},\check{g}\right]=0\quad.\label{eq:Usadel0}
\end{equation}
Here ${\bf h}=\mu_{B}{\bf B}$, ${\bm\sigma}=(\sigma^{x},\sigma^{y},\sigma^{z})$ and $\tau_{2,3}$
are Pauli matrices spanning spin and Nambu space, respectively, $\omega$ is the Matsubara
frequency, $\Delta$ is the superconducting order parameter, and SOC
enters via the covariant derivative $\tilde{\nabla}_{k}\check{g}=\partial_{k}\check{g}-i\left[\hat{{\cal A}}_{k},\check{g}\right]$,
where $\hat{{\cal A}}_{k}=\alpha\left(\delta_{kx}\sigma^{y}-\delta_{ky}\sigma^{x}\right)$,  summation over repeated indices is implied, and
$k=x,y$ \footnote{If the magnetic field has a component out-of-plane one should include in Eq. (\ref{eq:Usadel0}) the usual U(1) vector potential which leads to
orbital effects. Here we are interesting in the spin-magnetic response
and neglect orbital  terms.}. The quasiclassical GF $\check{g}$ in Eq. (\ref{eq:Usadel0})
is a 4$\times$4 matrix in the Nambu-spin space, which satisfies the normalization
condition $\check{g}^{2}=1$. In the absence of spin-dependent fields
it reads $\check{g}_{0}=(\omega/E)\tau_{3}+(\Delta/E)\tau_{2}$, where
$E=\sqrt{\omega^{2}+\Delta^{2}}$. It is easy to check by substitution
into Eq. (\ref{eq:Usadel0}), that in the absence of Zeeman field
$\check{g}_{0}$ is also the solution of the Usadel equation for arbitrary
$\hat{{\cal A}}_{k}$.

To compute the response to an external
magnetic field we  linearize Eq. (\ref{eq:Usadel0}) with respect to
$\bf h$ and write the solution as $\check{g}\approx\check{g}_{0}+\delta\check{g}$. It is convenient to define $\delta\check{g}\equiv i\check{g}_{0}\left[\tau_{3},\check{g}_{0}\right]\hat{Q}$, where  $\hat{Q}$ is a matrix in spin space that satisfies the following equation \footnote{In deriving Eq. (\ref{eq:lin_Usadel}) we {used the linearized
normalization condition $\check{g}_0\delta\check{g}+\delta\check{g}\check{g}_0=0$}.}: 
\begin{equation}
D\tilde{\nabla}_{k}^{2}\hat{Q}-2E\hat{Q}={\bf h}\cdot{\bm\sigma}\quad,\label{eq:lin_Usadel}
\end{equation}
The excess spin density $\delta S^{a}$ is then determined by
\footnote{{ The deviation from the Pauli response, $\delta{\bf S}={\bf S}-\chi_P{\bf B}$,
is determined  by $\delta S^{a}=-(i/2)\pi TN_{F}\sum_\omega{\rm Tr}[\tau_{3}\sigma^{a}\delta {g}]$, where $\chi_P=-2N_F\mu_B$, and  $N_F$ is the density of states at the Fermi level. In the  normal state $\delta{\bf S}=0$  and therefore the generated  magnetic moment is $M_{0}=-\mu_{B}\chi_P B$. In the superconducting state at zero temperature and zero SOC $\delta S=-\chi_P B$ and hence the total  magnetization is zero.}}: 
\begin{equation}
\delta S^{a}=-2\pi TN_{F}\sum_{\omega}{\rm Tr}_{\sigma}\frac{\Delta^{2}}{E^{2}}\left[\sigma^{a}\hat{Q}\right]\quad.\label{eq:magnetic_moment}
\end{equation}
For a homogeneous infinite 2D superconductor, the solution $\hat{Q}_{b}=Q_{b}^{a}\sigma^{a}$
of Eq. (\ref{eq:lin_Usadel}) reads 
\begin{eqnarray}
Q_{b}^{z} & =&- h^{z}\left[2E+8D\alpha^{2}\right]^{-1}\label{eq:Qbz}\\
Q_{b}^{x,y} & = & -h^{x,y}[2E+4D\alpha^{2}]^{-1}\; .\label{eq:Qbpara}
\end{eqnarray}
Equations (\ref{eq:magnetic_moment})-(\ref{eq:Qbpara}) reproduce
the bulk spin response of Rashba superconductor \cite{Gorkov2001,edelstein2008anomalous}, which is finite even at $T=0$ and depends on the direction of the applied field. 

\begin{figure}
\begin{minipage}[t]{1\columnwidth}%
\includegraphics[scale=0.3]{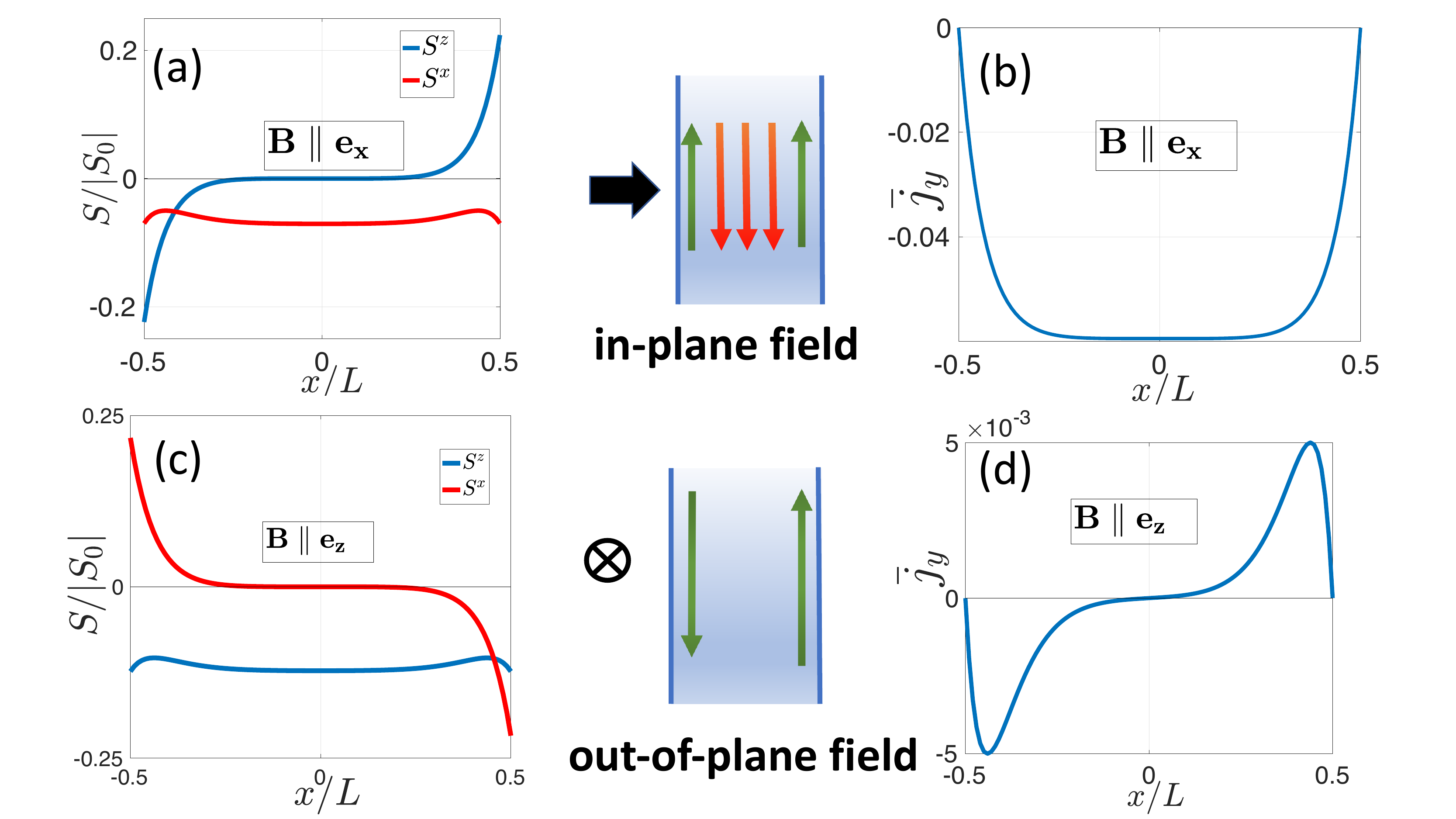}%
\end{minipage}
\caption{Spin density (panels a,c), and charge current density (panels b,d),  induced by an  in-plane (a-b) and out-of plane field (c-d),  { for $\alpha\xi_0=0.2$}, $L=10\xi_{0}$. The middle panels show schematically 
 the corresponding bulk (red arrows) and edge currents (green arrows) \label{fig:Anomalous-charge-current}}
\end{figure}

This situation changes drastically in a finite system. First, we  assume  that the system is infinite in $y$-direction, and bounded to the region $|x|<L/2$ in the $x$-direction. The solution to Eq. (\ref{eq:lin_Usadel}) can be written as the sum of the bulk contribution and a contribution from the sample edges,   $\hat{Q}=\hat{Q}_{b}+\delta\hat{Q}(x)$.  According to Eq. (\ref{eq:lin_Usadel}) the latter satisfies: 
\begin{eqnarray}
D\partial_{xx}^{2}\delta Q^{x}-\left(4D\alpha^{2}+2E\right)\delta Q^{x}+4D\alpha\partial_{x}\delta Q^{z} & = & 0\label{eq:Qx_eq}\\
D\partial_{xx}^{2}\delta Q^{z}-\left(8D\alpha^{2}+2E\right)\delta Q^{z}-4D\alpha\partial_{x}\delta Q^{x} & = & 0\;.
\label{eq:Qz_eq}
\end{eqnarray}
The last terms in these equations describe precession of the excess spin, caused by SOC. Importantly, the precession terms are finite only for inhomogeneous systems. The boundary conditions to the above equations are obtained by imposing zero-current at the edges, $x=\pm L/2$ \cite{Bergeret2013,Tokatly2017}: 
\begin{equation}
\left.\partial_{x}\delta\hat{Q}-i\alpha\left[\sigma^{y},\delta\hat{Q}\right]\right|_{x=\pm L/2}=i\alpha\left[\sigma^{y},\hat{Q}_{b}\right]\;.
\label{eq:BC_Q}
\end{equation}
Here the left hand side is proportional to the inhomogeneous spectral spin-current which cancels the bulk one in the right hand side. 
The boundary problem of Eqs. (\ref{eq:Qz_eq})-(\ref{eq:BC_Q}) has a nontrivial solution only if the right-hand-side in Eq. (\ref{eq:BC_Q}), that is the bulk spin-current, is finite. According to Eqs. (\ref{eq:Qbz})-(\ref{eq:Qbpara}), this is the case when the magnetic field has either $z$- or $x$-components. How to obtain the solution for $\delta Q^a$ is discussed in the Supplementary Material.  Here we present the spatial dependence of the induced spin  obtained from Eq. (\ref{eq:magnetic_moment}) and  plotted in  Figs.~\ref{fig:Anomalous-charge-current}(a,c). Both for in-plane (${\bf B}\parallel{\bf e}_x$), and for out-of-plane (${\bf B}\parallel{\bf e}_z$) fields, in addition to the longitudinal spin, a  transverse commponent of the  spin-density is generated. The latter is localized at the edges of
the sample with opposite sign on opposite sides { and decay into the bulk over the coherence length $\xi_s$.}  These results  generalize the theory of magnetic response for Rashba superconductors  \cite{Gorkov2001,edelstein2008anomalous} to the case of finite samples. 

In addition to the  finite spin response at $T=0$,  the SOC in superconductors  also leads to the  spin-galvanic effect, that is, a creation of charge currents by a Zeeman field \cite{yip2002two,dimitrova2007theory,edelstein2005magnetoelectric,Konschelle2015}. In the stripe geometry (see middle panels of  Fig.~\ref{fig:Anomalous-charge-current})
the so called anomalous charge current is induced in $y$-direction, { $j_{y}^{an}=(\theta/m)(\partial_{x}\delta S^{z}-2\alpha\delta S^{x})$ \cite{Tokatly2017}, where $\theta=2D\tau\alpha^{2}$ is a dimensionless parameter which in normal  systems describes the spin-charge conversion\cite{sanz2019non}. 
Within the diffusive approximation it is a small parameter which we treat perturbatively.}  Here $\delta S^{a}$ is obtained by substituting the solution of Eqs.~(\ref{eq:Qx_eq})-(\ref{eq:Qz_eq}) into Eq.~(\ref{eq:magnetic_moment}). This results in  
\begin{eqnarray}
j_{y}^{an}&=&-4\pi \theta \frac{TN_F}{m}\sum_{\omega}\frac{\Delta^{2}}{E^{2}}\left(\partial_{x}\delta Q^{z}-2\alpha\delta Q^{x}-2\alpha Q_{b}^{x}\right)\nonumber\\&\equiv&
j^{an}_{edge}(x)+j^{an}_{b}\; .
\label{eq:jy_bulk}
\end{eqnarray}
In the second line we identify two contributions to the anomalous current: the bulk contribution $j^{an}_{b}$, widely studied in homogeneous superconductors \cite{yip2002two,dimitrova2007theory,edelstein1995magnetoelectric,Konschelle2015}  and  given by the last term in the brackets in the first line and  (red arrows in middle panels in Fig.~\ref{fig:Anomalous-charge-current}), and the boundary contribution $j^{an}_{edge}$, determined by the first two terms. The latter are  localized at the edges of the sample within the scale of superconducting coherence length (green arrows in middle panels of Fig.~\ref{fig:Anomalous-charge-current}).   In the geometry under consideration,  the "bulk" contribution to the current is finite only for fields applied  across the stripe  ($x$-direction).

The  spatial dependence of the charge current density  is shown on Fig.~\ref{fig:Anomalous-charge-current} b and d for fields in $x$- and $z$-direction, respectively.  Because of zero spin-current condition,  Eq.~\eqref{eq:BC_Q}, the charge current of Eq.~\eqref{eq:jy_bulk} also vanishes at the boundaries.  When the  field is applied in $x$-direction, Fig. \ref{fig:Anomalous-charge-current}b,
both, the bulk and edge contributions, are finite. The maximal value of the  total current is the "bulk" value reached deeply inside the sample, away from the edges. The spatial distribution of the current is symmetric and the net current through the stripe is non-zero.
In contrast, if the field is applied  in $z$-direction, Fig.~\ref{fig:Anomalous-charge-current}d, there is no bulk contribution, 
because $Q_{b}^{z}$ does not contribute to the current, see Eq.~(\ref{eq:jy_bulk}). Only edge currents, opposite on opposite sides, appear.  Clearly in this case the total charge current vanishes. 

\begin{figure}
\begin{minipage}[t]{1\columnwidth}%
\includegraphics[scale=0.25]{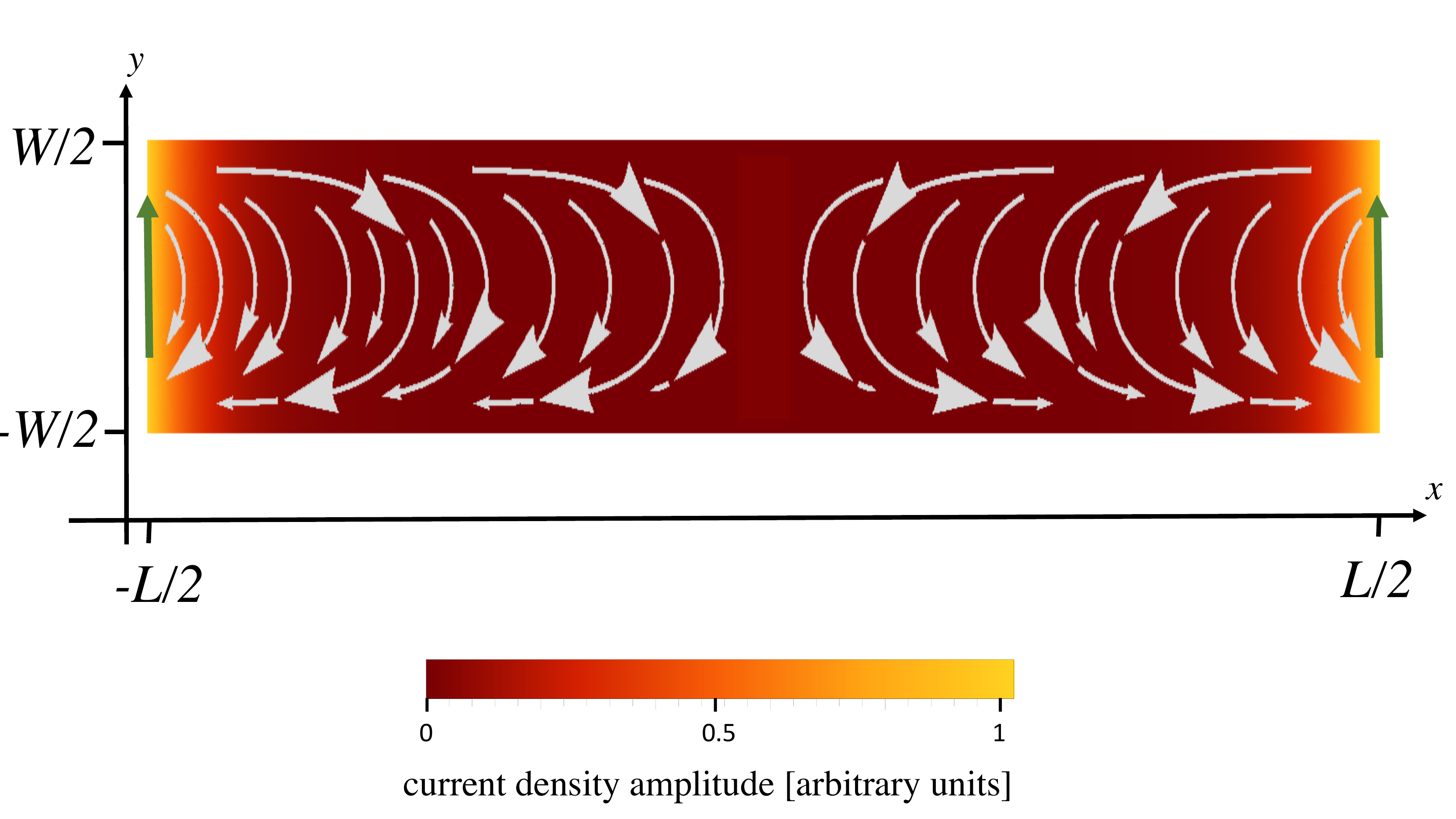}%
\end{minipage}
\caption{The current flow in a finite 2D superconductor with SOC when the field
is applied in $x$-direction. The green arrows represents the edge
contributions to the anomalous SH-current. The color-scale shows the
decay of the current amplitude away from the interface. \label{fig:currents}}
\end{figure}

The above results apply for an infinite superconducting stripe, and whether the obtained  currents may  exist in real finite systems depends on transverse boundary conditions. For example, if a finite stripe is wrapped in a cylinder, periodic boundary conditions indeed allow for the above current patterns when the field is applied along the  the cylinder axis \cite{dimitrova2007theory}.
Here we consider a more experimentally relevant situation: a finite  2D superconductor of rectangular shape,  which occupies
the region $|x|<L/2$ and $|y|<W/2$ (see Fig.~\ref{fig:currents}).  The charge current through all boundaries must vanish. For out-of-plane field this condition is trivially satisfied by closing the boundary streamlines, which generates a circulating edge current. More interesting is the case of in-plane field.  In this case the anomalous current has the same direction at both edges, Fig.~\ref{fig:Anomalous-charge-current}(b), and as sketched in  Fig.~\ref{fig:Anomalous-charge-current}(c)  one expects generation of closed streamlines at each edge. 

{  Specifically, the total charge current in the superconductor reads, 
\begin{equation}
{\bf j}=\delta(z)\left[\frac{e n_{s}}{2m}\left(\nabla \varphi-\frac{e}{c}{\bf A}\right)+ j^{an}(x)\hat{\bf y}\right]\; ,
\label{eq:tot_current}
\end{equation}
where $n_{s}$ is the superfluid density in the 2D strip. The superconducting phase $\varphi(x,y)$
and the vector potential ${\bf A}({\bf r})$ are determined, respectivelly, by the continuity equation and the Maxwell equation,  
\begin{equation}
\nabla\cdot{\bf j}=0 \quad\text{and}\quad \nabla^2{\bf A}=-\frac{4\pi}{c}{\bf j} \; ,\label{eq:Maxwell}
\end{equation}
which should be solved with the zero-current condition at the edges, ${\bf n}\cdot{\bf j}|_{edge}=0$, where ${\bf n}$ is a unit vector normal to the edges of the sample. 
We assume that $n_s$ is homogeneous within the strip. Then, by choosing the gauge with $\nabla\cdot{\bf A}=0$, the continuity equation is reduced to the 2D Laplace equation for the phase, $\nabla^2\varphi=0$.

In the problem defined by Eqs.~(\ref{eq:tot_current})-(\ref{eq:Maxwell}) one identifies three length scales: (i) a mesoscopic scale of the order of the coherence length $\xi_s$, over which the anomalous current $j^{an}(x)$ decays away from the edges, (ii) the Pearl length $\Lambda=mc^2/4\pi e^2n_s$ that is the scale controlling Meissner effect in 2D superconductors, and  (iii) the sample geometry scales $W$, $L$. In the following we consider the typical situation when $\xi_s\ll \Lambda,W,L$ and analyze the current distribution in a narrow strip with $W\ll L$. In this case the 
anamolaous current in Eq.~(\ref{eq:tot_current}) can be written as $j^{an}(x)=I_{y}^{edge}\big[\delta(x+L/2)+\delta(x-L/2)\big]$, and the currents near opposite edges at $x=\pm L/2$ can be treated independently. 

The current streamlines are sketeched in Fig.~\ref{fig:currents}.  Whereas the anomalous current is strongly localized at the edges (green arrows  in  Fig.~\ref{fig:currents}),  the counterflow supercurrent compensating the anomalous one, decays over a macroscopic scale determined by the width of the sample and/or the Pearl length $\Lambda$. If $W\ll \Lambda$, one can neglect ${\bf A}$ in Eq.~(\ref{eq:tot_current}) and the problem can be solved using the procedure described in Refs.~\cite{borge2019boundary,sanz2019non}. In this limit the counterflow decays exponentially over the scale $W$, and sufficiently far from the edges it takes the form, 
\begin{equation}
 \label{eq:curr_asymp-1}
 (j_x,\, j_y)  \propto I_y^{edge}
 \frac{e^{-\frac{\pi|x\pm L/2|}{2W}}}{W}\left(\sin\frac{\pi y}{2W},\, \cos\frac{\pi y}{2W}\right)
\end{equation}

In the opposite limit of $W\gg\Lambda$ one can neglect the corner effects and apply the method of images and conformal mapping \cite{norris,zeldov} to compute screening supercurrents induced by an external current filament at the edge of 2D superconducting half-plane. This gives a power-law asymptotic decay of the counterflow supercurrent,  
\begin{equation}
 \label{eq:curr_asymp-2}
 j_y \propto I_y^{edge}\sqrt{\Lambda}|x\pm L/2|^{-3/2}.
\end{equation}

The total current generates a finite orbital angular momentum ${\cal L}_{z}$ at each edge, see Fig. \ref{fig:Spin-density}c, which is computed from the general definition ${\cal L}_{z}=m\int\int dxdy(xj_{y}-yj_{x})/2$ and  Eqs. (\ref{eq:curr_asymp-1}) and  (\ref{eq:curr_asymp-2})
\begin{equation}
%{\cal L}_{z}=-m\frac{W^{2}}{4}I_{y}^{edge}\int\frac{dk}{2\pi}\frac{\tanh^{2}k}{k^{2}}\approx-m\frac{W^{2}}{8}I_{y}^{edge}\;.
{\cal L}_{z}\propto-m f(W)I_{y}^{edge}\;,
\label{eq:Lz}
\end{equation}
where $f(W)=W^2$ in the limit $W\ll\Lambda$ and $f(W)=W^{3/2}\sqrt{\Lambda}$ when  $W\gg\Lambda$ \cite{Note6}.
The  total magnetic moment is 
given by ${\cal M}_{z}=\mu_{B}({{\cal L}_{z}}/{\hbar}+{{\cal S}_{z}})$, 
where $\mu_{B}$ is the Bohr magneton \footnote{The spin magnetic moment is given by ${\bf \mu_{spin}}=(-g|e|/2m)(\hbar/2){\bf \sigma}=\mu_{B}\sigma$}. The  total spin angular momentum accumulated at the edge is obtained  by integrating the  $z$-component of the  spin,  Eq.~(\ref{eq:magnetic_moment}), ${\cal S}_{z}=\int_{-L/2}^0\delta S^z(x)dx$.
Analytical expressions for both spin and orbital angular momenta at $T=0$ can be found in two  cases \cite{NoteSM}  }
{
\begin{equation}
{\cal S}_{z}  \propto -N_F h_x W\xi_0
\begin{cases}
      \xi_0\alpha, & \text{for  $\xi_0\alpha\ll 1$}\\
     (\alpha\xi_0)^{-3}, & \text{for $\xi_0\alpha\gg 1$}\; ,
  \end{cases}
     \label{eq:Sz_final_T0}\\
\end{equation}
}and 
\begin{equation}
{\cal L}_{z}  \propto -N_Fh_x\theta f(W)
\begin{cases}
\xi_0\alpha, & \text{for  $\xi_0\alpha\ll 1$}\\
(\xi_0\alpha)^{-4}\ln (\xi_0\alpha), & \text{for $\xi_0\alpha\gg 1$}
\end{cases}
\label{eq:Lz_final_T0}\;
\end{equation}
with $\xi_0=\sqrt{D/2\Delta}$.  Both contributions have the same sign.  The spin angular momentum scales with $W$, while  ${\cal L}_{z}$ scales with $W^{2}$ or $W^{3/2}$ depending on the ratio $W/\Lambda$,  and therefore dominates in macroscopic samples.

In conclusion, we  present the theory of the magnetic response of finite size Rashba  superconductors.  When the field is applied in-plane,  on the one hand, a finite out-of-plane spin polarization localized at the edge of the sample on the scale of superconducting coherence length appears. On the other hand,  the SOC also leads to supercurrents circulating in the sample. Both the spin and the orbital momentum of supercurrents contribute to the total magnetic moment, which is induced at the edges and can be measured by state-of-the-art magnetic sensors\cite{granata2016nano,maletinsky2012robust}.  Whereas the contribution from the spin angular momentum scales with the width  $W$ of a rectangular stripe, the contribution from the currents scales with $W^\gamma$, with $\gamma>1$ and therefore dominates in large samples.      
There are several superconducting materials with  Rashba SOC  in  which our findings can be verified. These range  from Pb and Tl-Pb monolayers \cite{qin2009superconductivity,sekihara2013two,brun2014remarkable,matetskiy2015two}, to thin MoS$_2$, NbRe, $\beta$-Bi$_2$Pd films   \cite{yuan2014possible,cirillo2016superconducting,lv2017experimental}, and 2D superconductivity at the  LaAlO$_3$/SrTiO$_3$ interface \cite{dikin2011coexistence,bert2011direct,kalisky2012critical,hurand2015field}.  
{ A particular  interesting system  has been studied recently \cite{moodera2020}.
It consists of EuS grown on top of Au (111) surface which is proximitized by an adjacent superconductor.
According to our theory, the exchange field induced by EuS, together with the large Rashba SOC in the Au 2D interface band, should lead to the transverse edge magnetization and edge supercurrents even in the absence of an external applied field.}

{\it Acknowledgements.-} We acknowledge funding by the Spanish Ministerio de Ciencia, Innovaci\'on y Universidades (MICINN) (Projects No.  FIS2016-79464-P and No. FIS2017-82804-P), by Grupos Consolidados UPV/EHU del Gobierno Vasco (Grant No. IT1249-19), and by EU's Horizon 2020 research and innovation program under Grant Agreement No. 800923 (SUPERTED).

\bibliographystyle{apsrev4-1}
%\bibliography{library}

\end{document}